# Influence of Morphology on Blinking Mechanisms and Excitonic Fine Structure of Single Colloidal Nanoplatelets

Zhongjian Hu[1,†], Ajay Singh[1,†], Serguei V. Goupalov[2,3], Jennifer A. Hollingsworth[1] & Han Htoon[1,*]

[1]Center for Integrated Nanotechnologies, Material Physics and Applications Division

Los Alamos National Laboratory, Los Alamos, NM 87545

[2]Department of Physics, Jackson State University, Jackson, MS 39217

[3]Ioffe Institute, St. Petersburg 194021, Russia

[†]Equal contribution

[*]Correspondence and requests for materials should be addressed to H.H.

 (email: htoon@lanl.gov)

**Abstract**

Colloidal semiconductor nanoplatelets (NPLs) with electronic structure as quantum wells have recently emerged as exciting materials for optoelectronic applications. Here we investigate how morphology affects important photoluminescence (PL) properties of single CdSe and core/shell CdSe/CdZnS nanoplatelets. By analyzing PL intensity-lifetime correlation and second-order photon correlation results, we demonstrate that, irrespective of morphology, Auger recombination cannot be responsible for PL blinking of single NPLs. We propose that hot carrier trapping plays a significant role in blinking and find that a rough shell induces additional nonradiative channels presumably related to defects or traps of an imperfect shell. Polarization-resolved PL spectroscopy analysis reveals exciton fine-structure splitting on the order of several tens of meV in rough-shell NPLs at room temperature, which is attributed to exciton localization and substantiated with theoretical calculations taking into account the NPL shape and electron-hole exchange interaction.

Colloidal semiconductor nanocrystals (NCs) have been attracting considerable interest in both fundamental and applied scientific research in the past two decades.[1-3]. Compared with the extensively studied 0D quantum dots (QDs) and 1D nanorods/nanowires, only in the past several years have colloidal 2D nanoplatelets (NPLs) with controllable morphology and dimensions been introduced.[4-9] The NPLs usually have a lateral dimension larger than the bulk exciton Bohr radius and a thickness of several monolayers (MLs), thus representing free-standing quasi 2D quantum wells. With the thickness that is well-defined and can be engineered with atomic precision, the NPLs have narrow photoluminescence (PL) linewidths at room temperature not affected by inhomogeneous broadening.[7, 10-12] In addition, the quasi 1D carrier confinement in NPLs results in a giant oscillator strength and a fast recombination of band-edge excitons.[10, 12, 13] These remarkable and unique photophysical properties have made the colloidal NPLs very attractive for a broad range of applications such as lasing and light emitting diodes.[14-19]

PL blinking (i.e., intensity intermittency), a phenomenon that has been extensively studied for colloidal QDs, has also been reported for single NPLs.[12, 20, 21] Several studies have attributed the blinking of single NPLs to Auger recombination.[12, 21] On the other hand, for 2D quantum wells, the electronic confinement occurs only in one dimension, thereby imposing a theoretically stringent momentum conservation rule for Auger recombination.[15] In fact, suppressed Auger recombination in NPLs has been demonstrated in multiple works recently.[15, 22, 23] Such contradictory reports warrant further investigations regarding the PL blinking mechanism and Auger processes in NPLs. In addition, 2D NPLs have a strong quantum confinement in the thickness direction and a large surface area. Therefore, shell growth on core NPLs and shell morphology are expected to have a strong impact on the excited state photophysics and PL spectral properties. Although some fundamental PL properties of core/shell NPL structures have been reported by several studies,[20, [24, 25] up to now morphological effects on PL blinking behavior (in connection with its relation to Auger recombination), exciton fine-structure, and PL polarization have not been investigated.

Herein, we study in detail impacts of CdZnS shell growth and shell morphology—smooth or rough—on PL properties of single CdSe NPLs in terms of PL blinking, Auger recombination, biexciton quantum yield, PL spectrum, polarization, and exciton fine-structure. We reveal that, irrespective of shell growth and morphology, Auger recombination plays little role in PL blinking of single NPLs and propose that hot carrier trapping plays a significant role. A rough CdZnS shell results in blinking that is associated with lifetime changes, implying opening of additional nonradiative channels presumably related to defects or traps of an imperfect shell. We further report differences in low-temperature PL spectral linewidth and emission polarization as a function of shell morphology. In particular, polarization-resolved spectral measurements reveal exciton fine-structure splitting in the presence of a rough shell, which is attributed to exciton localization.

## Results

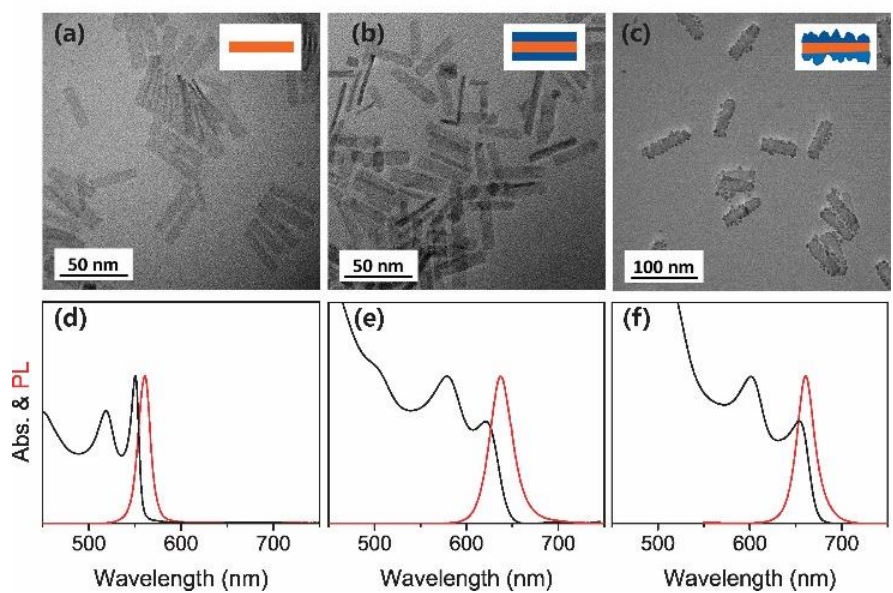

**Figure 1 | Electron microscopy and ensemble solution spectra.** Transmission electron microscopy (TEM) images of core CdSe NPLs (**a**), CdSe NPLs with smooth CdZnS shell (**b**), and CdSe NPLs with rough CdZnS shell (**c**). The insets in (**a-c**) display schematics of side-view of NPLs. The bottom panels (**d-f**) show the absorption (black) and PL (red) spectra of solutions for the core NPLs, the smooth-shell NPLs, and the rough-shell NPLs, respectively.

**Structural and ensemble optical characterization.** The CdSe NPLs were synthesized following previously reported method.[10] CdZnS shell were added to the core CdSe NPLs using modifications of two literature methods, i.e., layer-by-layer growth[20] and one-pot continuous shell growth[25], resulting in core/shell (C/S) NPLs with smooth and rough shell morphology, respectively (see Methods and Supplementary Note 1). Fig. 1(a-c) displays TEM images for the core CdSe NPLs, the smooth CdZnS shell NPLs, and the rough-shell NPLs, respectively. The core NPLs have a lateral dimension of ~7 ± 1 nm in width and ~50 ± 7 nm in length. The thickness of the core NPLs is ~1.5 nm, which corresponds to 5 CdSe monolayers, based on analysis of TEM images and the known relationship between thickness and the absorption excitonic peak.[24] For both the smooth- and rough-shell NPLs, the lateral dimensions remain similar to those of the NPL core, indicating that shell growth occurred mainly on the top and bottom sides. The shell thickness is ~1 nm on both sides. The smooth-shell NPLs exhibit uniform contrast, indicative of the smooth shell morphology (Fig. 1b), while the rough-shell NPLs exhibit significant contrast variation (Fig. 1c), indicative of island-like structures with an average diameter of 3-5 nm. These are not simply continuous grain boundaries, as the shell height is observed to fluctuate across the shell surface. In this way, the shell thickness of ~1 nm is only an average for the rough-shell NPLs.

Absorption and PL spectra of NPLs suspended in hexanes/chloroform solutions are shown in Fig. 1(d-f). For the core NPLs, two absorption peaks are observed at 545 nm and 515 nm that correspond to the heavy- and light-hole excitons, respectively. Emission peaks at ~555 nm with a full-width-at-half-maximum (FWHM) of 12 nm (48 meV). The absorption and emission spectral results are consistent with the previously reported data for 5-monolayer CdSe NPLs.[12, 24] CdZnS shell growth induces strong red shifts in both absorption and emission, which can be attributed to the electronic wave function extension from the CdSe core into the shell material.[20] The smooth-shell NPLs and the rough-shell NPLs have emission peaks at 645 nm (quantum yield: 60-65%) and 665 nm (quantum yield: 45-50%), respectively. This difference in the emission maximum between the smooth- and rough-shell NPLs is likely due to the rough shell being slightly thicker, or to the observed thickness fluctuations. In addition, compared to the core NPLs, the C/S NPLs have broader emission linewidths: ~60 meV and ~57 meV for the smooth- and rough-shell NPLs, respectively. This observation is also similar to the results reported by Tessier et al. and are attributed to the extension of electron-phonon coupling into the shell material and the high phonon energy of the shell material.[20, 23]

**Blinking of single NPLs.** PL intensity blinking (i.e., intermittency) is a ubiquitous phenomenon observed in individual semiconductor nanocrystals.[26] The blinking behaviors have also been observed for NPLs.[12, 20] Analogous to the cases of the C/S heterostructures of QDs, the growth of shell greatly reduces PL blinking. This is reflected in the time traces of fluorescence intensity shown in Fig. 2(a-c): the C/S NPLs exhibit a much longer "on" time compared with the core NPLs. Such an improvement is further demonstrated in a statistical analysis of the on-time fraction of ~50 NPLs under one-hour continuous excitation (Supplementary Fig. S1).

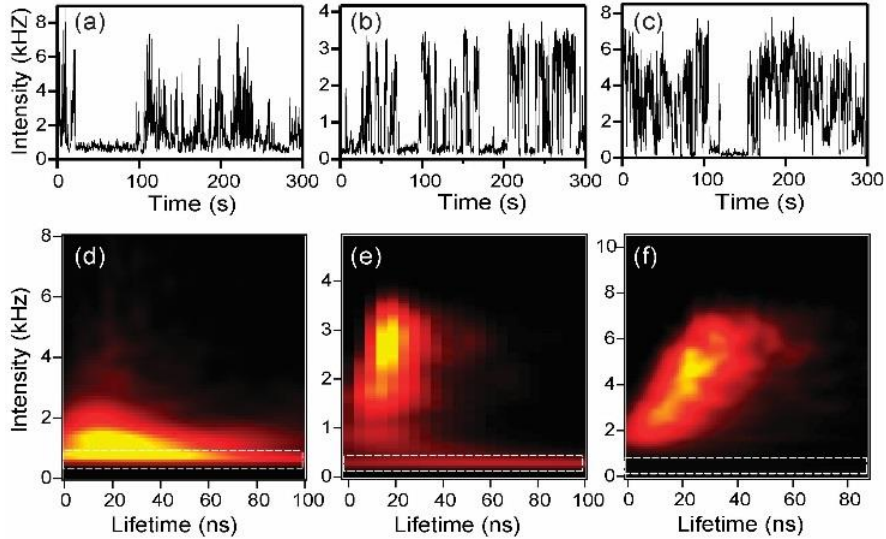

**Figure 2 | PL intensity fluctuation and correlated lifetime-intensity distribution.** Representative PL intensity time traces of the core CdSe NPLs (**a**), the smooth-shell CdSe/CdZnS NPLs (**b**), and the rough-shell CdSe/CdZnS NPLs (**c**). Panel (**d-f**) display the FLID diagrams for the corresponding time traces (**a-c**) for three types of NPLs, respectively. The white dashed box in (**d-e**) indicates the signal from the background.

The phenomenon of PL blinking, since its first report by Nirmal et al in single QDs, has been intensively studied in terms of the fundamental mechanisms.[26, 27] Although the deterministic mechanism of PL blinking is still under debate, the widely-accepted explanation of the off-state (or the grey state that weakly fluoresces) is that the non-radiative Auger recombination as a result of photo-induced charging is responsible. To understand the in-depth mechanism of the blinking behavior in the NPLs, we analyze the PL lifetime for photons collected at different intensity levels and demonstrate correlations between PL intensity and lifetime in fluorescence lifetime-intensity distribution (FLID) 2D diagrams.[28] Fig. 2(d-f) displays the FLID diagrams for PL intensity time traces shown in Fig. 2(a-c), respectively. For the core NPL and the smooth-shell NPL, as shown in Fig. 2(d) and 2(e) the PL lifetime does not exhibit significant change with the PL intensity fluctuation. Detailed comparisons of PL lifetimes of several PL intensity bands for the NPLs can be found in Supplementary Fig. S2. The blinking behavior observed in the core and the smooth-shell NPLs excludes Auger recombination mechanism. Rather, it reminds us of the B-type blinking phenomenon of the CdSe/CdS QDs with medium shell thickness as reported in our previous work.[28] For the B-type blinking, it is proposed that there exist nonradiative decay channels that can intercept hot electrons, which are sufficiently energetic to tunnel to the surface trap sites after photoexcitation, before they relax to the lowest band edge states. In this case, the PL intensity is proportional to the number of the electrons relaxed to the band edge states. However, since the relaxation dynamics from the band edge states to the ground state are not disturbed by the hot electron trapping, the PL lifetime remains unchanged. Considering the significantly large surface area of the NPLs and the shell thickness of only several monolayers, we think the blinking behavior observed in the core and the smooth-shell NPLs can also be explained by such a hot carrier trapping mechanism.

By contrast, for the rough-shell NPLs, PL lifetime varies along with PL intensity as shown in Fig. 2c and 2f. Such a blinking phenomenon is referred to as A-type blinking in our previous work[28] and rationalized by non-radiative Auger recombination. Here the recombination of charged excitons results in the off-state, wherein the recombining electron-hole pair transfers its energy to the third charge nearby, while the recombination of neutral band edge excitons leads to the PL emission of on-state. Some previous studies have tentatively attributed the blinking behavior observed for NPLs to Auger recombination mechanism.[12, 21] However, there have been several reports showing that Auger process in NPLs is greatly hampered.[15, 22, 23, 29, 30] Additionally, electronic confinement occurs only in one dimension in 2D quantum wells, thereby imposing a theoretically strict momentum conservation rule for Auger recombination.[15] These contradictory reports surrounding Auger recombination warrant an in-depth investigation regarding its role in the blinking behavior of NPLs.

**Auger recombination suppression and biexciton.** Auger recombination is closely related to the biexciton emission in nanocrystals. In order to understand these two important characteristics and elucidate the blinking mechanism in the NPLs, we perform the second-order photon correlation, $g^{(2)}$, analysis. For typical $g^{(2)}$ traces obtained with pulsed laser excitation, the area ratio between the center peak (delay time at 0) to the side peaks, $R$, signifies the emission nature of the emitters. An $R$ value of 0, i.e., photon antibunching, is characteristic of single quantum emitters, while a non-zero $R$ value indicates either emitter clustering or biexciton emission.

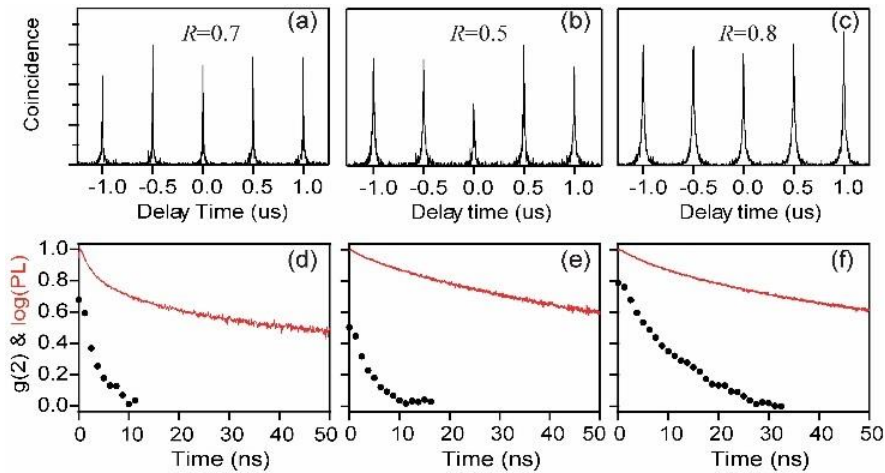

**Figure 3 | Single-photon correlation experiments.** Photon correlation data of the core CdSe NPL (**a**), the smooth-shell NPL (**b**), and the rough-shell NPL (**c**). The corresponding PL decays (red) and the decays of $R$ value as a function of gated time (black dotted) are shown in (**d-f**). These data were obtained from the same respective NPLs demonstrated in Fig. 2.

Fig. 3(a-c) demonstrates $g^{(2)}$ plots and corresponding $R$ values, as insets, in each panel for the same three NPLs, for which the blinking traces are shown in Fig. 2. One can see that all three types of NPLs exhibit high $R$ values above 0.5. To disentangle the effects of NPLs clustering and biexciton emission, we employ the time-gated antibunching technique developed by our group.[31] In this software-based time gating approach, the contribution of the possible faster biexciton emission can be reduced (removed) by applying a time-gate that is approaching (longer than) the lifetime of biexcitons. Therefore, a quick decay of the *R* value to zero with the gated delay time signifies that the R value is contributed by the biexciton emission of a single NPL. Fig. 3(d-f) displays decays of the *R* value as functions of the applied gated delay time for the three NPLs. One can see that the *R* value drops much faster than the overall fluorescence, which is shown as a red curve for comparison, for all three types of NPLs. The fast decay of *R* towards 0 within a gated delay time less than 10-30 ns suggests that the initial *R* values above 0.5 are caused by high biexciton quantum yield ($Q_{BX}$) in the NPLs rather than NPLs clustering. For isolated nanocrystals, *R* is directly proportional to the ratio of biexciton quantum yield to single exciton quantum yield ($Q_X$), that is, $R = g^{(2)}(0)/g^{(2)}(t) = Q_{BX}/Q_X$.[32,33] Our data are consistent with the recent results reported by Ma et al.[21] The efficient biexciton yield in the NPLs implies a greatly suppressed biexciton Auger recombination. In addition, according to the statistical scaling, the biexciton Auger recombination efficiency provides an upper limit for the charged exciton Auger decay efficiency.[34-36] Therefore, Auger recombination by charged excitons is also suppressed.

On the basis of the above discussions, we can conclude that charged excitons cannot be responsible for blinking behavior in the rough NPLs. We would attribute the faster decay of lower intensity levels to some non-radiative channels that are related to imperfect shell morphology, such as surface defects, traps, etc. This result is consistent with the observations made on ensembles of CdSe/CdS NPLs by Kunneman et al, who reported hole trapping at a defect site basing on transient absorption and terahertz conductivity measurements.[22] In addition, a close look at the ratios of PL intensity levels and their corresponding lifetimes reveals that the intensity drops faster than the PL lifetime. For instance, as shown in Supplementary Fig. S2c and S2f, the ratio between the highest to the lowest highlighted intensity bands is ~4.3, while the ratio between their corresponding lifetimes is only ~2.3. This observation indicates that, besides the nonradiative processes related to rough morphology, the hot carrier trapping mechanism also contributes to the blinking of rough-shell NPLs. Our results regarding the blinking of NPLs not only provide a detailed understanding of the blinking mechanism and its relation with morphology, but indicate possible routes to making non-blinking NPLs by structure engineering. For instance, a thicker smooth shell is desirable, as it would reduce both the number of surface defects/traps and the hot carrier tunneling probability.

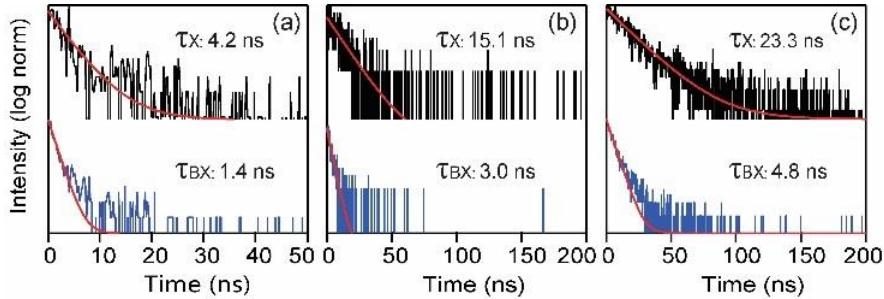

**Figure 4 | PL lifetimes of exciton and biexciton.** Extracted PL decays and monoexponential fits (red) of biexciton (blue) and single exciton (black), based on the post-selection of photons corresponding to the zero time delay peak in the photon correlation data (Fig. 3a-c), for the core CdSe NPL (**a**), the smooth-shell NPL (**b**), and the rough-shell NPL (**c**).

Our analysis of about twenty NPLs for each type reveals average $R$ values of 0.63±0.28, 0.61±0.14, 0.78±0.14 for the core NPLs, the smooth-shell NPLs, and the rough-shell NPLs, respectively. It can be seen that the shell deposition and morphology do not significantly alter the biexciton yield. We further extract the biexciton and single exciton decay lifetimes using an approach of post-selection of photons on the basis of photon correlation data reported by Canneson et al.[37] Fig. 4 demonstrates the corresponding biexciton and single exciton decays overlaid with monoexponential fits for the three representative NPLs shown in Fig. 2 and Fig. 3. The analysis of about 20 NPLs show that the core NPLs have average lifetimes of 1.2±0.2 ns for biexciton emission and 3.1±0.8 ns for single exciton emission, while the rough-shell NPLs have the values of 3.8±1.2 ns and 17.8±5.8 ns, and the smooth-shell NPLs have the values of 3.2±0.8 ns and 17.3±3.6 ns, respectively. Compared to the core NPL, the C/S NPLs exhibit longer exciton and biexciton lifetimes. This can be ascribed to the electron wavefunction delocalization into the shell material. Furthermore, the rough-shell NPLs have slightly longer biexciton lifetimes, which is in line with the observation that a longer gated delay time is usually required to remove all biexciton contributions (Fig. 3f). The longer biexciton lifetime observed for the rough-shell NPLs could be explained by localized excitons in the rough shell that may take longer to reach the core and contribute to biexcitonic emission. This also may result from greater electronic wavefunction extension into an on-average thicker shell (resulting from the rough shell morphology).

**Photoluminescence spectra.** We further collect PL spectra of single NPLs in order to identify the effects of shell growth and shell morphology on the spectral properties. Fig. 5(a-c) exhibit the typical single NPL spectra collected at room temperature and low temperature (10 K). The linewidth of room-temperature emission spectra is 45±2, 57±4, 53±2 meV for the core, the smooth-shell, and the rough-shell NPLs, respectively. These values are close to the PL linewidth results of the ensemble solutions (Fig. 2), suggesting negligible inhomogeneous broadening irrespective of the type and the morphology of NPLs. At room temperature, there is no detectable difference in both linewidth and spectral structure between the smooth-shell NPLs and the rough-shell NPLs. Going from room temperature to 10 K, the emission blue shifts and the linewidth becomes much narrower for all types of NPLs. The blue shifted emission has been reported by Tessier et al and explained with the Varshni relation between the band gap and temperature, which predicts an increased band gap with decreasing temperature.[12, 20] Interestingly, we notice that the smooth-shell NPLs typically exhibit a slightly narrower linewidth of PL emission (~9±1 meV) compared to the rough-shell NPLs (~12±2 meV). In addition, the emission spectra of rough-shell NPLs usually exhibit a more pronounced red tail emission. Tessier et al have also observed a similar tail emission in low-temperature PL spectra (analogous to the shape of the black spectrum in Fig. 5b) for CdSe/CdZnS NPLs, yet irrespective of the shell morphology, and attributed it to low energy surface traps related to defects in the shell or at nanocrystal surface.[20] Compared to the results reported by Tessier et al, the more pronounced tail emission in our case (Fig. 5c) may be due to more rough shell morphology in our rough-shell NPLs. We think that the subtle differences in both tail emission and linewidth for the two types of C/S NPLs in our case could be due to the variation in shell morphology. That is, the rough-shell NPLs have slightly higher density of localized trap states as a result of rough shell morphology.[38, 39]

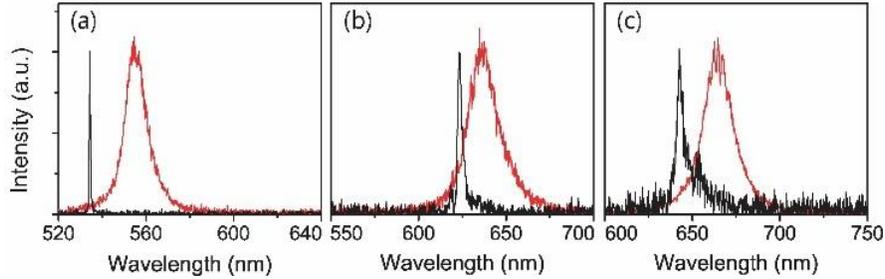

**Figure 5 | PL spectra of single NPLs.** Representative PL spectra at room-temperature (red) and 10 K (black) of the core CdSe NPLs (**a**), the smooth-shell NPLs (**b**), and the rough-shell NPLs (**c**).

**Emission polarization anisotropy and polarization-resolved PL spectra of single NPLs.** Polarization characteristics of NCs convey information on the nature of the transition dipole and exciton fine structure.[40-42] We first investigate the polarization anisotropy of total PL emission by splitting PL image into two orthogonal polarizations (i.e., horizontally/vertically polarized emission $I_H/I_V$) using a Wollaston prism.[41, 42] The polarization is modulated by rotating a half-wave plate in front of the prism. Fig. 6a demonstrates the normalized horizontal emission, $I_{NH}$ ($I_{NH}=I_H/(I_H+I_V)$), as a function of the emission angle. A fit with a sine-squared function to the data reveals the degree of polarization, $P=(I_{NH,max}-I_{NH,min})/(I_{NH,max}+I_{NH,min})$. The core NPLs exhibit high degree of polarization with P values of 0.68±0.20. The NPLs with shell, irrespective of the shell morphology, exhibit lower degrees of polarization with P values of 0.44±0.11, 0.42±0.13 for the smooth and the rough-shell NPLs, respectively (Supplementary Fig. S3). Next, we acquire the polarization-resolved PL spectra of individual NPLs by spectrally dispersing the two orthogonal PL channels after the Wollaston prism using imaging spectrometer.[41, 42] With this detection scheme, horizontal- (H) and vertical-polarized (V) PL components from individual NPLs are observed as pairs of spectra in the upper and lower regions of the CCD camera. Fig. 6b-6d displays orthogonally polarized PL emission spectra (black and red curves) as a function of emission detection angle for a core NPL (Fig. 6b), a smooth-shell NPL (Fig. 6c), and a rough-shell NPL (Fig. 6d), respectively. The polarization-resolved PL spectra of the core NPLs exhibit high degree of linear polarization with no spectral splitting between two orthogonally polarized PL spectra, suggesting that the PL emission is originated from a linear transition dipole. Surprisingly, polarization-resolved PL spectra of ~66% of the rough-shell NPLs (Fig. 6d) and ~45% of smooth-shell NPLs reveal three energetically distinct, linearly polarized emissive states with energy splitting of ~25 meV. Out of these three states, the highest and lowest energy states have almost identical polarization orientation whereas the polarization of the middle state exhibits ~55° phase shift relative to that of the highest and lowest energy states (Fig. 6d). The remainder of the smooth-shell (~55%, Fig. 6c) and rough-shell NPLs (~34%) show broad PL spectra, without an obvious three-peak-splitting feature and with low degree of linear polarization consistent with total PL polarization anisotropy (Fig. 6a).

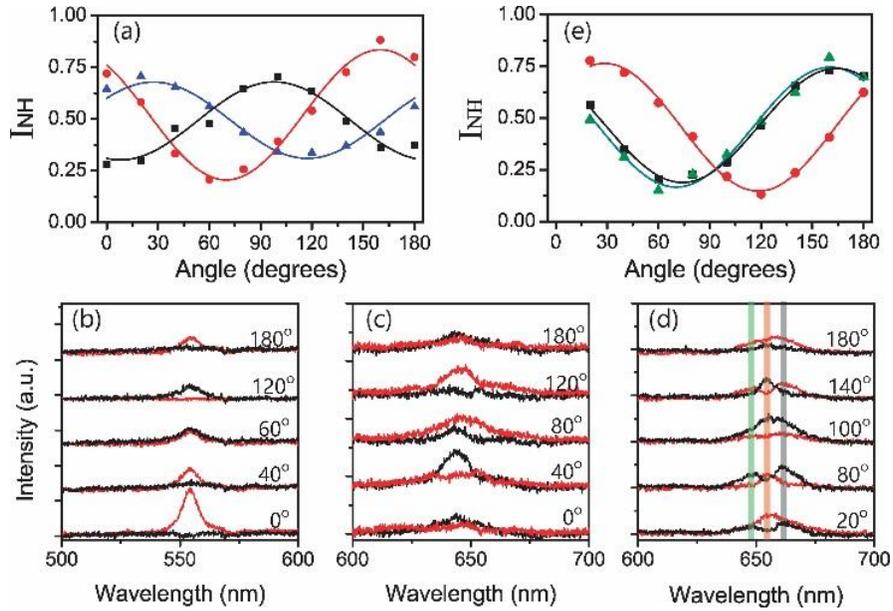

**Figure 6 | Polarization experiments of single NPLs.** (**a**) Typical polarization modulations obtained from imaging as a function of emission collection angle for single NPLs: the core NPLs (red), the smooth-shell NPLs (black), and the rough-shell NPLs (blue), respectively. The solid lines are fits to the data. (**b-d**) Polarization-resolved spectra collected at different emission detection angles (shown as inset) for a single core NPL (**b**), a single smooth-shell NPL (**c**), and a single rough-shell NPL (**d**), respectively. Red and black curves correspond to horizontal- (H) and vertical-polarized (V) spectral components. (**e**) The polarization modulations measured for the three highlighted 2-nm-wide wavelength bands in panel (**d**) shown with the same color. The solid lines are fits to the data.

Recent studies of relatively symmetric NPLs reported observation of polarization behavior consistent with a 2 D planar dipole corresponding to the heavy-hole exciton radiative doublet.[43, 44] Because our NPLs are highly asymmetric with the width and length of ~7 nm and ~50 nm, respectively, we expect the emissive exciton state to be split into two states and the low energy state of the two to dominate the emission with linear transition dipole oriented along the long axis of the NPL. While this picture can explain strongly linearly polarized emission of the core only NPLs, it cannot explain the emergence of the three emissive states in cases of the smooth- and rough-shell NPLs. Because the three emissive states were observed mostly in rough-shell NPLs, we hypothe-size that the localization of the excitons in confinement potential minima resulting from the thickness fluctuations of NPLs is responsible. To test this hypothesis, we construct a theoretical model of a NPL that localizes an exciton in a region of the width $2L_y$=70 Å in the *y*-direction and varying localization length along the *x*-direction ($2L_x$), which has the size of the surface roughness (Fig. 7a). For such a NPL, we expect four states, with two orthonormal states each for delocalized exciton ($E_{3x}$ & $E_{4y}$,) and localized exciton ($E_{1x}$ & $E_{2y}$), respectively (Fig. 7a). To estimate energy splitting between these states, we use the well-known expression for the exciton resonance frequency renormalization due to the long-range electron-hole exchange interaction, $\delta\omega_0^{(\alpha)}$, for excitons localized in quantum wells[45, 46] (the index $\alpha$ indicates the polarization of the exciton sublevels) and assume that the envelope wave function of an exciton localized on the roughness has a Gaussian form.[46] Detailed description of theoretical calculations can be found in Supplementary Note 3. Since the NPLs sizes are less than the emission wavelength, the final exchange-induced splitting can be expressed as

$$\delta\omega_0^{(y)} - \delta\omega_0^{(x)} \approx -\frac{\omega_{LT} a_B^3 L_x L_y}{2\pi \tilde{a}^2} \int_0^\infty dt\, t^2 \int_0^{2\pi} d\varphi \cos 2\varphi\, e^{-t^2 \cos^2\varphi L_x^2/2} e^{-t^2 \sin^2\varphi L_y^2/2}$$
$$\times \int dz\, \Phi(z) \int dz'\, \Phi(z') e^{-t|z-z'|}$$

where, $\omega_{LT}$ and $a_B$ are the longitudinal-transverse splitting and the Bohr radius of a free exciton in the bulk material, respectively, $\tilde{a}$ is the effective two-dimensional Bohr radius, $\Phi(z)=\varphi_{e1}(z)\varphi_{hh1}(z)$ is the product of the electron and hole envelope wave functions along the axis of the quantum well or NPL (here we use the envelope wave functions for carriers confined in a quantum well with infinitely high barriers). For our estimate we will take $\hbar\omega_{LT}$=0.95 meV,[47] $a_B$=56 Å. Nanoplatelet excitons are known to have the effective two-dimensional Bohr radius < 10 Å,[48] for an estimate we take $\tilde{a}$=7 Å. The resulting dependence of the anisotropic exchange splitting on the size of the localizing potential along $x$, $L_x$, at the fixed size along $y$, $L_y$=35 Å, is shown in Fig. 7b. The splitting is zero for $L_x$=$L_y$. Comparing the observed splitting with this dependence one can estimate the size of the localizing potential as $2L_x$. Our calculation yields a splitting between two delocalized exciton states $E_{4y}$-$E_{3x}$ to be 63.5 meV as the delocalized exciton corresponds to the limit $L_x \to \infty$ (in particular, $L_x$>200 Å). Because this splitting is larger than the thermal energy at room temperature, we can assume that the highest energy state $E_{4y}$ is not accessible to the exciton population and therefore leading to the appearance of three peaks. Localization of exciton in ~100 Å region (i.e. $L_x$ ~ 50 Å) gives rise to $E_{2y}$-$E_{1x}$ and $E_{3x}$-$E_{2y}$ splittings of ~25 meV, as observed in our experiments. $E_{1x}$ and $E_{3x}$ are linearly polarized along the long axis of the NPLs and $E_{2y}$ is polarized in the perpendicular direction. This corresponds to our observation that the two side peaks have essentially the same polarization orientation.

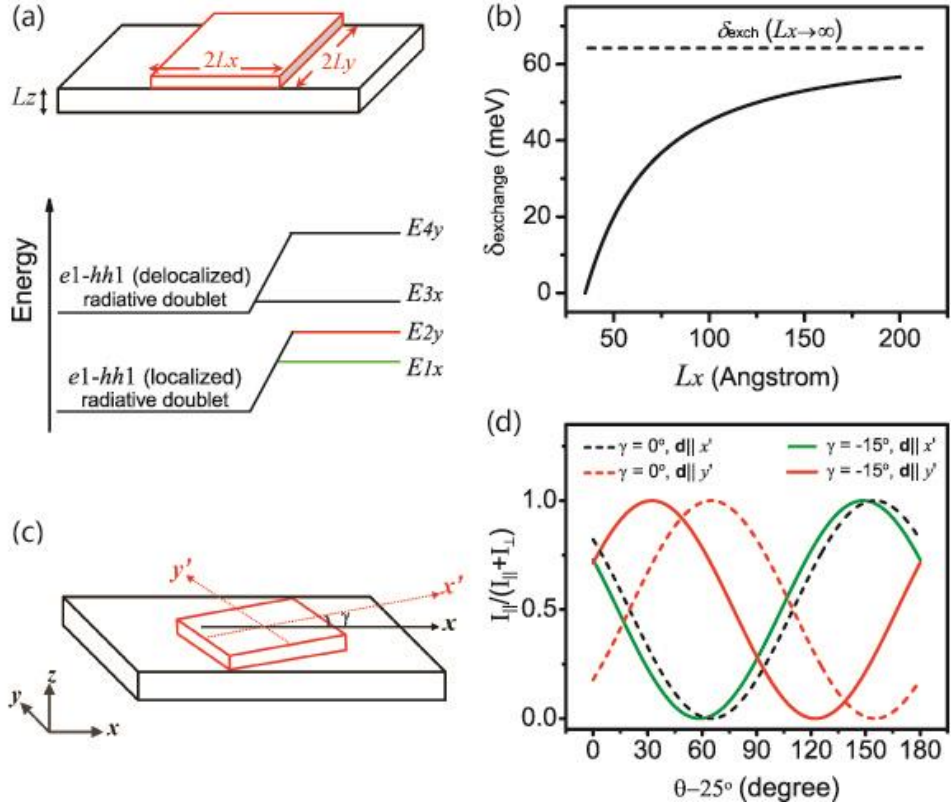

**Figure 7 | Theoretical modeling and calculations.** (**a**) Top: A schematic of a NPL with a width of $2L_y$ and a thickness of $L_z$ (black box). The theoretical modeling assumes that the exciton is localized on the red box, which has a width of $2L_y$, a length of $2L_x$, and a thickness equivalent to the surface roughness. Bottom: Fine structural scheme of energy levels of the exciton radiative doublets for both delocalized and localized excitons before and after electron-hole exchange interaction is taken into account, respectively. (**b**) Dependence of the anisotropic exchange splitting, $\hbar(\delta\omega_0^{(y)}-\delta\omega_0^{(x)})$, on the localization size along $x$, $L_x$. The localization size along $y$-axis is fixed: $2L_y=70$ Å. (**c**) A possible orientation of the exciton localization region that can result in a polarization "nonorthogonality". (**d**) Angular dependencies of $I_\parallel^{x'}(\theta)/(I_\parallel^{x'}+I_\perp^{x'})$ and $I_\parallel^{y'}(\theta)/(I_\parallel^{y'}+I_\perp^{y'})$ (i.e., polarization modulation of emission dipole along the $x'$ and $y'$ axis, respectively) for $\gamma=0°$ and $\gamma=-15°$. $\theta$ represents the angle between the $x$-axis and the "horizontal" polarization of the analyzer.

While the middle peak in the above theoretical picture should be polarized normal to the two side peaks, the result shown in Fig. 6e shows that the polarization of the middle peak only has a ~55° phase shift. This inconsistency could result from a misalignment of the orientation of the exciton localization region (due to surface roughness fluctuation) with respect to the main axes of the NPL (Fig. 7c, also refer to Supplementary Note 3 for details). To illustrate this point, we construct a model as displayed in Fig. 7c, for which we assume that the main axes of the exciton localization region ($x'$, $y'$) form a small angle $\gamma$ with respect to the main axes of the NPL ($x$, $y$). A drastic difference in high-frequency dielectric constants between the NPLs ($\varepsilon_{in}$~6) and the ambient air ($\varepsilon_{out}$~1) leads to a strong depolarization field in the $y$-direction perpendicular to the axis of the NPL elongation. As a result of misalignment, the dipole oriented along $y'$ has a small projection onto $x$ axis (cf. Fig. 7c). But, since the electric field along $y$ axis is suppressed due to the depolarization effect, the electric field of light, emitted by the dipole polarized along $y'$, has comparable components along $x$ and $y$. Similarly, the dipole oriented along $x'$ has a small projection onto $y$ axis. But the depolarization effect leads to the fact that the corresponding small component of the electric field in the emitted light is further suppressed. As a result, the depolarization effect strongly affects polarization of the light emitted by the dipole polarized along *y'* and has almost no effect on polarization of the light emitted by the dipole polarized along *x'*. Our detailed calculation (see Supplementary Note 3) reveals that a misalignment angle of $\gamma \sim 15°$ can approximately reproduce the large phase shift in the polarized PL intensity from *y'* polarized dipole and the negligible shift in the polarized PL intensity from *x'* polarized dipole (Fig. 7d) as we observed in Fig. 6e. Taken together, the theoretical results are in excellent agreement with our experimental observations.

## Discussion

We have systematically investigated the morphological effects on PL properties of CdSe/CdZnS NPLs. The deposition of CdZnS shell on core CdSe NPLs not only reduces PL blinking of single NPLs but, more importantly, results in variations in the optical and electronic properties depending on shell morphology.

We have shown that, regardless of shell growth and morphology, Auger recombination cannot be responsible for PL blinking of single NPLs. Instead, we propose that hot carrier trapping plays a significant role in blinking. We have found that the deposition of a rough CdZnS shell leads to opening of additional nonradiative channels and slightly broader and less symmetric PL spectra compared to smooth-shell NPLs, which are possibly due to the existence of trap or defect sites in an imperfect shell.

We have performed polarization-resolved spectral analysis of PL from single NPLs and found that single NPLs exhibit polarized emission whereas shell deposition reduces the degree of polarization. Exciton fine-structure splitting on the order of tens of meV has been observed in room-temperature PL spectra of rough-shell NPLs. We have shown that exciton can get localized at a surface irregularity resulting from the rough shell morphology. Both the fine-structure splitting and PL polarization resulting from exciton localization are sensitive to the size and shape of the exciton localization region. The PL polarization is further affected by the depolarization effect determined by the shape of NPL. These conclusions are supported by theoretical calculations which are in excellent agreement with experimental observations.

**Methods**

**Nanoplatelets Synthesis.** CdSe NPLs with emission around 555 nm are synthesized according to previously published procedure with slight modifications.[10] Smooth CdZnS shell on CdSe NPLs were synthesized following atomic layer deposition approach with some modification.[20] To achieve rough CdZnS, we used previously published continuous shell growth protocol with minor modification.[25] The extra ligands in the nanoplatelets solutions were removed using ethanol for precipitation and re-dispersing in hexane or chloroform. Details about the materials and synthesis of NPLs are available in Supplementary Note 1.

**Single NPLs spectroscopy.** Single NPLs samples were prepared by drop-casting suspensions of NPLs in hexanes onto clean glass coverslips. Single NPLs spectroscopy experiments were carried out on a typical laser scanning confocal optical microscope using a pulsed 403 nm diode laser (1 MHz, ~70 ps pulse width). PL emission from individual NPLs was collected with a 100×, 0.85 NA objective lens and cleaned with long pass filters. Time-resolved PL and $g^{(2)}$ experiments were performed with two single photon avalanche photodiodes (SPAD, SPCM-AQR-14, PerkinElmer) placed in a Hanbury-Brown-Twiss geometry and connected to a Hydraharp 400 TCSPC system operating in a time-tag time-resolved mode. Single NPLs PL spectra were measured by directing PL emission to a spectrograph (SP-2300i, Princeton Instruments) and imaged with a CCD camera. Emission polarization anisotropy and polarization-resolved PL experiments were performed in a wide-field mode. Herein, the 403 nm diode laser was focused to a spot size with diameter of ~60 μm and was used to excite the samples. The PL images and spectra were collected with the same spectrometer and CCD camera as used in the confocal mode. A Wollaston prism was used to spatially split the emission into its respective horizontal and vertical locations along the spectrometer slit. This enabled a simultaneous measurement of two orthogonal components. A half-wave plate was deployed in front of the Wollaston prism to tune the emission angle.

**Acknowledgments**

action equal opportunity employer, is operated by Los Alamos National Security, LLC, for the National Nuclear Security Administration of the U.S. DOE under contract DE-AC52-06NA25396. Z.H., J.A.H. and H.H. were supported by LANL Laboratory Directed Research and Development Funds. A.S. was supported by a Los Alamos National Laboratory Director's Postdoctoral Fellowship. S.V.G. would like to thank M.O. Nestoklon and E.L. Ivchenko for useful discussions. The work of SVG was supported by the National Science Foundation (NSF-CREST Grant HRD-1547754).

**Author contributions**

Z.H. performed and analyzed the spectroscopy experiments under the guidance of H.H. A.S. synthesized the materials. S.V.G. performed the simulations. H.H. and J.A.H. conceived and planed the project. Z.H., S.V.G., and H.H. wrote the paper with input from other authors.